\documentclass[letterpaper,american,reprint, aps, prb, superscriptaddress]{revtex4-1}
\usepackage[T2A,T1]{fontenc}
\usepackage[latin9]{inputenc}
\usepackage{units}
\usepackage{array}
\usepackage{gensymb}
\usepackage{array,booktabs}
\usepackage{textcomp}
\usepackage{amsmath}
\usepackage{textgreek}
\usepackage{amssymb}
\usepackage{graphicx}
\usepackage{dcolumn}
\usepackage{bm}

\providecommand{\tabularnewline}{\\}

\makeatletter
\DeclareRobustCommand{\greektext}{%
  \fontencoding{LGR}\selectfont\def\encodingdefault{LGR}}
\DeclareRobustCommand{\textgreek}[1]{\leavevmode{\greektext #1}}
\ProvideTextCommand{\~}{LGR}[1]{\char126#1}

\DeclareRobustCommand{\cyrtext}{%
  \fontencoding{T2A}\selectfont\def\encodingdefault{T2A}}
\DeclareRobustCommand{\textcyr}[1]{\leavevmode{\cyrtext #1}}

\pdfpageheight\paperheight
\pdfpagewidth\paperwidth

\DeclareRobustCommand{\cyrtext}{%
  \fontencoding{T2A}\selectfont\def\encodingdefault{T2A}}
\DeclareRobustCommand{\textcyr}[1]{\leavevmode{\cyrtext #1}}

\makeatother

\usepackage{babel}
\begin{document}

\title{Bulk properties of van-der-Waals hard ferromagnet VI$_{3}$}

\author{Suhan Son}

\affiliation{Center for Correlated Electron Systems, Institute for Basic Science,
Seoul 08826, Republic of Korea}

\affiliation{Department of Physics and Astronomy, Seoul National University, Seoul
08826, Republic of Korea}

\author{Matthew J. Coak}

\affiliation{Center for Correlated Electron Systems, Institute for Basic Science,
Seoul 08826, Republic of Korea}

\affiliation{Department of Physics and Astronomy, Seoul National University, Seoul
08826, Republic of Korea}

\affiliation{Cavendish Laboratory, Cambridge University, J.J. Thomson Ave, Cambridge
CB3 0HE, UK}

\author{Nahyun Lee}

\affiliation{Center for Correlated Electron Systems, Institute for Basic Science,
Seoul 08826, Republic of Korea}

\author{Jonghyeon Kim}

\affiliation{Department of Physics, Yonsei University, Seoul 03722, Republic of
Korea}

\author{Tae Yun Kim}

\affiliation{Department of Physics and Astronomy, Seoul National University, Seoul
08826, Republic of Korea}

\affiliation{Center for Theoretical Physics, Seoul National University, Seoul
08826, Republic of Korea}

\author{Hayrullo Hamidov}

\affiliation{Cavendish Laboratory, Cambridge University, J.J. Thomson Ave, Cambridge
CB3 0HE, UK}

\address{Navoiy Branch of the Academy of Sciences of Uzbekistan, Galaba Avenue,
Navoiy, Uzbekistan}

\affiliation{National University of Science and Technology \textquotedblleft MISiS\textquotedblright ,
Leninsky Prospekt 4, Moscow 119049, Russia}

\author{Hwanbeom Cho}

\affiliation{Center for Correlated Electron Systems, Institute for Basic Science,
Seoul 08826, Republic of Korea}

\affiliation{Department of Physics and Astronomy, Seoul National University, Seoul
08826, Republic of Korea}

\author{Cheng Liu}

\affiliation{Cavendish Laboratory, Cambridge University, J.J. Thomson Ave, Cambridge
CB3 0HE, UK}

\author{David M. Jarvis}

\affiliation{Cavendish Laboratory, Cambridge University, J.J. Thomson Ave, Cambridge
CB3 0HE, UK}

\author{Philip A.C. Brown}

\affiliation{Cavendish Laboratory, Cambridge University, J.J. Thomson Ave, Cambridge
CB3 0HE, UK}

\author{Jae Hoon Kim}

\affiliation{Department of Physics, Yonsei University, Seoul 03722, Republic of
Korea}

\author{Cheol-Hwan Park}

\affiliation{Department of Physics and Astronomy, Seoul National University, Seoul
08826, Republic of Korea}

\affiliation{Center for Theoretical Physics, Seoul National University, Seoul
08826, Republic of Korea}

\author{Daniel I. Khomskii}

\affiliation{II. Physikalisches Institut, Universit{\"a}t zu K{\"o}ln D-50937 K{\"o}ln Germany}

\author{Siddharth S. Saxena}

\affiliation{Cavendish Laboratory, Cambridge University, J.J. Thomson Ave, Cambridge
CB3 0HE, UK}

\affiliation{National University of Science and Technology \textquotedblleft MISiS\textquotedblright ,
Leninsky Prospekt 4, Moscow 119049, Russia}

\author{Je-Geun Park}

\affiliation{Center for Correlated Electron Systems, Institute for Basic Science,
Seoul 08826, Republic of Korea}

\affiliation{Department of Physics and Astronomy, Seoul National University, Seoul
08826, Republic of Korea}

\date{\today}
\begin{abstract}
We present comprehensive measurements of the structural, magnetic
and electronic properties of layered van-der-Waals ferromagnet VI$_{3}$
down to low temperatures. Despite belonging to a well-studied family
of transition metal trihalides, this material has received very little
attention. We outline, from high-resolution powder x-ray diffraction
measurements, a corrected room-temperature crystal structure to that
previously proposed and uncover a structural transition at 79~K, also seen in the heat capacity.
Magnetization measurements confirm VI$_{3}$ to be a hard ferromagnet
(9.1~kOe coercive field at 2~K) with a high degree of anisotropy,
and the pressure dependence of the magnetic properties provide evidence
for the two-dimensional nature of the magnetic order. Optical and
electrical transport measurements show this material to be an insulator
with an optical band gap of 0.67~eV - the previous theoretical predictions
of $d$-band metallicity then lead us to believe VI$_{3}$ to be a
correlated Mott insulator. Our latest band structure calculations
support this picture and show good agreement with the experimental
data. We suggest VI$_{3}$ to host great potential in the thriving
field of low-dimensional magnetism and functional materials, together
with opportunities to study and make use of low-dimensional Mott physics.
\end{abstract}
\maketitle

Two-dimensional van-der-Waals (vdW) magnetic materials have in recent
years become the subject of a wide range of intense research \citep{Ajayan2016}.
While a large portion of research into two-dimensional materials has
centered on graphene, the addition of magnetism into such a system
leads to many interesting fundamental questions and opportunities
for device applications \citep{Park2016,Kuo2016,Zhou2016,Samarth2017,Burch2018}.
Particularly for future spintronics applications, semiconducting or
metallic materials which exhibit ferromagnetism down to monolayer
thickness are an essential ingredient. This has led to a large volume
of recent publications on two-dimensional honeycomb ferromagnet CrI$_{3}$
\citep{McGuire2015,Zhang2015,Wang2016,Huang2017,Lado2017,Klein2018}.
CrI$_{3}$ and VI$_{3}$ belong to a wider family of MX$_{3}$ transition
metal trihalides, with X = Cl, Br, I, which were synthesized in the
60s \citep{Juza1969,Dillon1965} but have since seen little interest
until recently \citep{McGuire2017}.

VI$_{3}$ is an insulating two-dimensional ferromagnet with a Curie
Temperature, T$\mathrm{_{c}}$, given as 55~K and reported to have
the layered crystal structure of BiI$_{3}$ with space group R-3 \citep{Trotter1966, Handy1950, Wilson1987}.
As shown in a recent review \citep{McGuire2017}, there is very little
available information on VI$_{3}$ other than the structure and the
expected $S=1$ from the $3d^{2}$ configuration of the vanadium sites.
Calculations using density functional theory, which additionally yield
the exchange constants, have suggested VI$_{3}$ to not only remain
ferromagnetic down to a single crystalline layer, but to also exhibit
Dirac half-metallicity, of interest for spintronic applications \citep{He2016}.

In these vdW materials, hydrostatic pressure forms an extremely powerful
tuning parameter. Given the weak mechanical forces between the crystal
planes, the application of pressure will dominantly have the effect
of pressing the \emph{ab} planes together, and gradually and controllably
pushing the system from two- to three-dimensionality. Additionally,
first-principles calculations have suggested in-plane strain and compression
to stabilize both antiferromagnetic phases and spin reorientation
in CrI$_{3}$ \citep{Zheng2017} and both VBr$_{3}$ and VCl$_{3}$
are antiferromagnetic \citep{Wilson1987}. 

Finding the correct crystal structure is crucial for accurate ab-initio
calculations of a new material's behavior and for general insight
into its properties. The structure of VI$_{3}$ has not been previously
outlined beyond its basic crystallographic family \citep{Trotter1966,Berry1969,Juza1969,Wilson1987}.
This earlier work assigned VI$_{3}$ as belonging to the BiI$_{3}$
type $R-3$ structure, but no detailed crystallographic results are
available or any description of temperature dependence. Here we present
the crystal structure of VI$_{3}$ from refined powder x-ray diffraction
data and find a transition to an alternative structure at low temperature.

\begin{figure}
\centering{} \includegraphics[width=1\columnwidth]{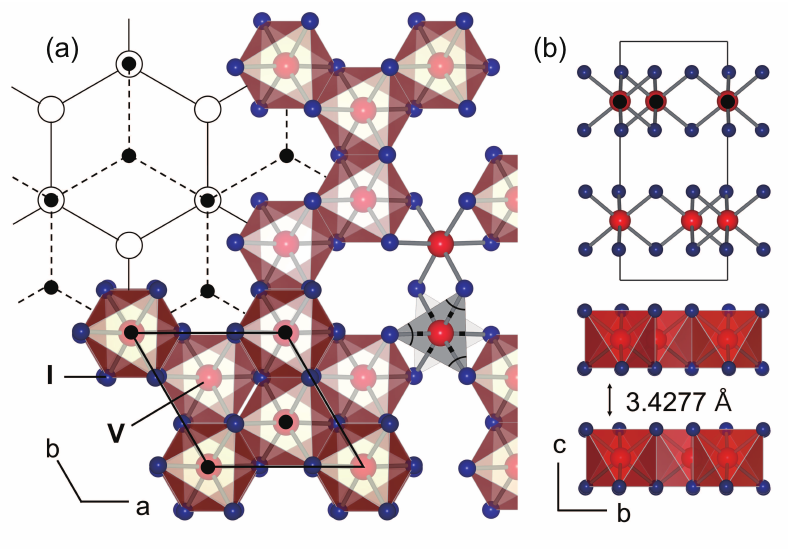}\caption{\label{fig:CrystalStructures}Views of the room-temperature crystal
structure of VI$_{3}$ along (a) the $c$ axis and (b) the $a$ axis.
Additional projections are shown in the Figs 8 and 9. }
\end{figure}

The results show good agreement between the high-resolution x-ray
diffraction (HR-XRD) measurement and refined data (Figs 6-9 and Tab. 1. The FULLPROF, GSAS-II and VESTA software suites were employed \citep{Rodriguez1993, Toby2013, Momma2011}). Contrary to the previous reports, we find the room
temperature structure to be fitted better with the space group of
$P-31c$ than $R-3$. Even though both structures show very similar
simulation results, several peaks were missing from the $R-3$ simulations,
and peak shapes and ratios are better fit by the $P-31c$ (Fig 7). The
$P-31c$ structure at room temperature is shown in Fig. \ref{fig:CrystalStructures}.
The unit cell has an ideal honeycomb bilayer of V formed of {[}VI$_{6}${]}$^{3-}$
octahedra, separated by a clear van-der-Waals gap. All the shortest
V-V and V-I bonds have the same lengths. 

\begin{figure}
\centering{}\includegraphics[width=1\columnwidth]{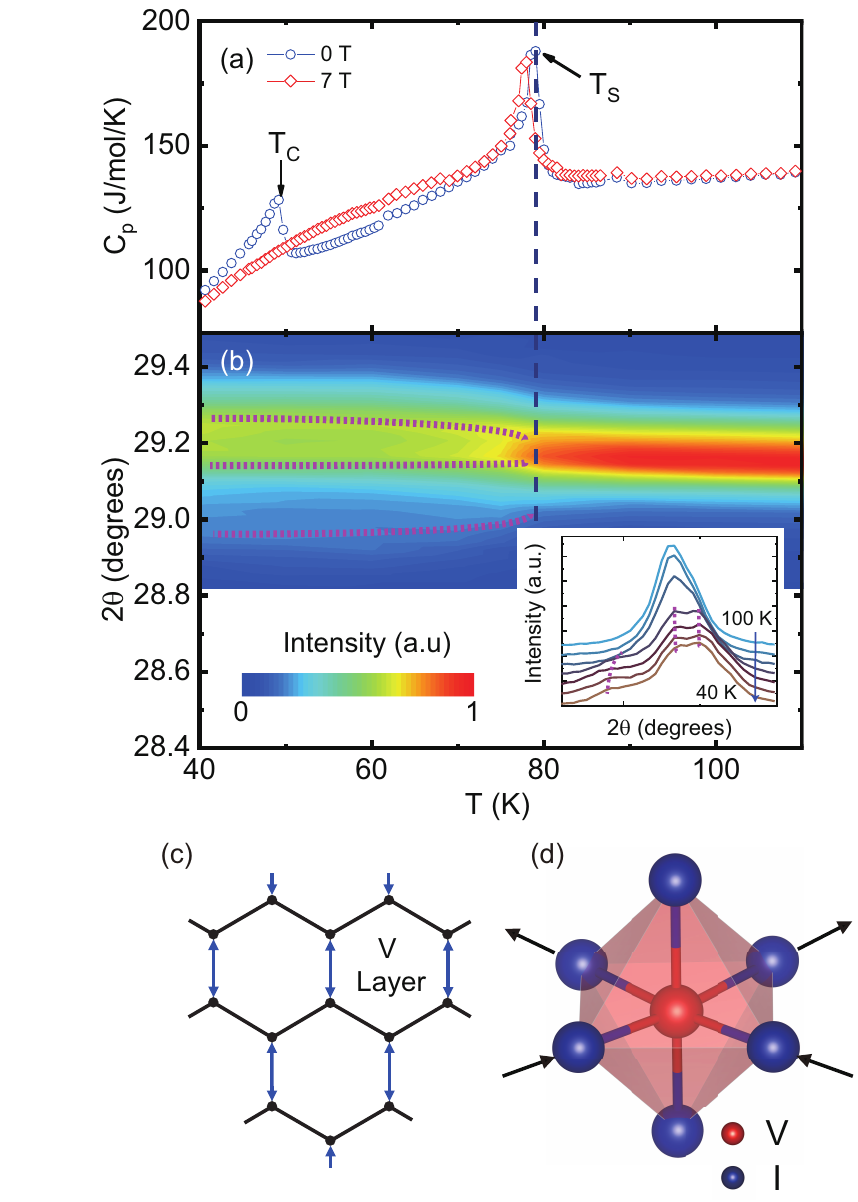} \caption{\label{fig:HeatCapAndStruuctureDetail}(a) Heat capacity of a VI$_{3}$
single crystal, in 0 and 7~T fields. The structural transition at
T$\mathrm{_{s}}$= 79~K is unaffected by magnetic field, unlike the
ferromagnetic transition at T$\mathrm{_{c}}$ = 50~K. (b) Detail
of temperature\textendash dependent x-ray powder diffraction patterns
showing peak splitting at T$\mathrm{_{s}}$. (c) and (d) Illustrations
of the anti-dimerisation of the V sites and the distortion
of the VI$_{6}$ octahedra in the low-temperature phase.}
\end{figure}

Following the powder diffraction patterns down to low temperature
additionally reveals a phase transition to a new low-temperature structure
below T$_{\mathrm{s}}$ = 79.0(5)~K. Fig. \ref{fig:HeatCapAndStruuctureDetail}
(a) and (b) show heat capacity and details of the x-ray diffraction
patterns as a function of temperature. A clear peak at 79~K in the
heat capacity and the peak splitting in the XRD patterns demonstrate
the existence of a sharp structural phase transition at this temperature. 

As one can see from the long debates about the low temperature structure
of $\alpha$-TiCl$_{3}$ \citep{Natta1961}, the overlap of many peaks
via symmetry breaking at low temperature gives multiple potential
solutions to the structure. Among the possible candidates, we obtained
the best results for a $C2/c$ structure, a sub-space group of the
room temperature structure. Contrary to MoCl$_{3}$ \citep{Hillebrecht1997}
and $\alpha$-TiCl3\citep{McGuire2017} undergoing dimerization of
the transition metal ions at low temperature, the V atoms here stretch
in a single direction from their perfect honeycomb - anti-dimerization
(see Fig. 2(c)). The short bonding length is 3.92(5)~\r{A} and the long
4.12(11)~\r{A}. Furthermore, accompanying the anti-dimerization of V
honeycomb atoms, the VI$_{6}$ octahedra experience an off-center distortion when the low-temperature structure is entered. Fig. \ref{fig:HeatCapAndStruuctureDetail}(d)
illustrates the structural deformation of the VI$_{6}$ octahedra at low
temperature. Plots of the refinement results and structural parameters
at 40 and 300 K are given in the Figs 6 and 7, Tab. 1. 

The peak seen in the heat capacity at 50~K (T$\mathrm{_{c}}$) we
identify as the ferromagnetic transition from the broadening of peak
width in an applied 7~T of magnetic field. All measurements were
reproducible over multiple thermal cycles, which rules out the effect
of sample degradation on the results. Like other transition-metal
halides with vdW gaps such as CrI$_{3}$ and $\alpha$-RuCl$_{3}$,
VI$_{3}$ is easily exfoliated by the common Scotch tape method \citep{Novoselov2004}
(Fig 11), placing it as a valuable system for studying two-dimensional
physics and applications.

\begin{figure}
\centering{} \includegraphics[width=1\columnwidth]{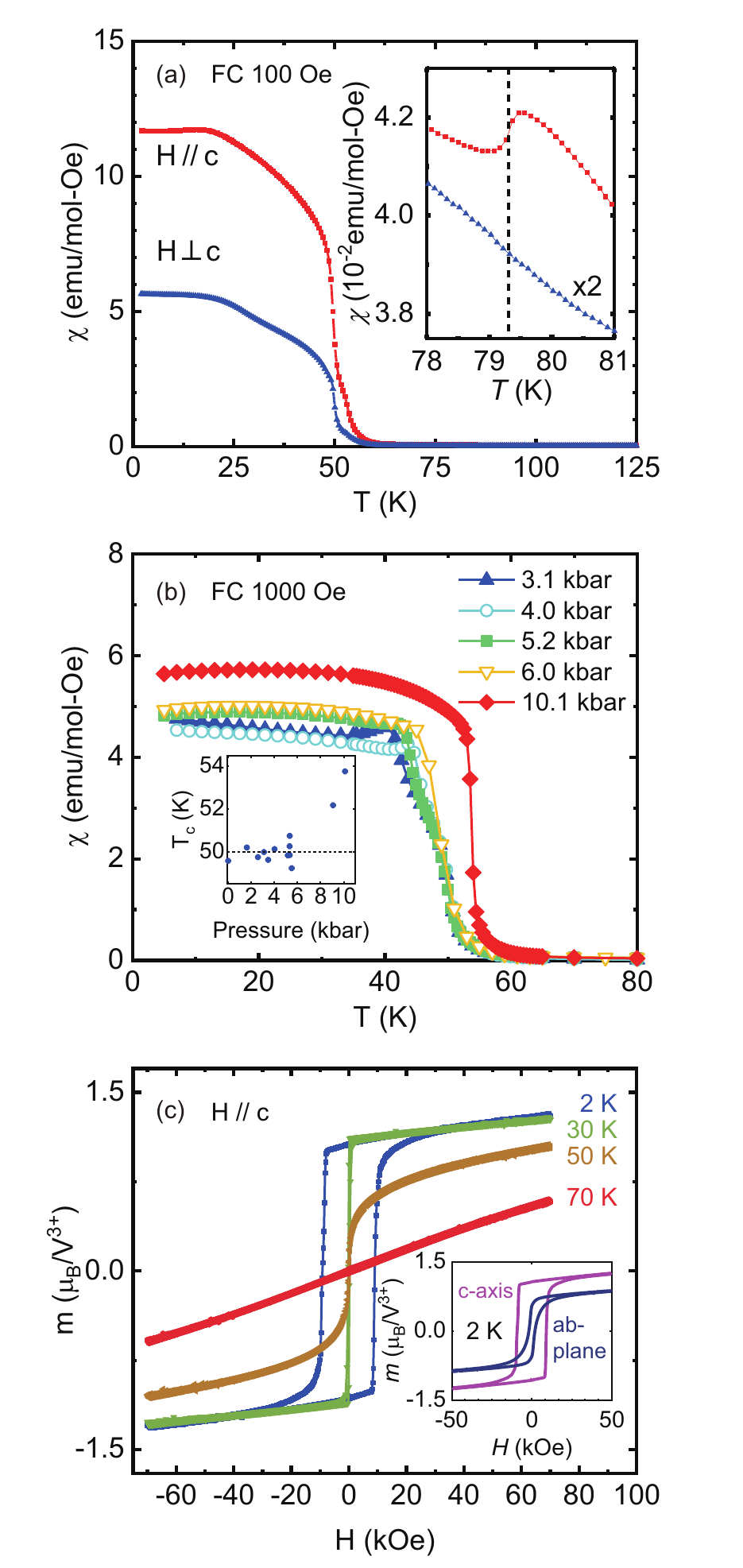}\caption{\label{fig:MagnetisationPlots}(a) Field-cooled magnetic susceptibility
of VI$_{3}$ with applied 100~Oe field parallel and perpendicular
to the crystallographic $c$ axis. Inset shows detail around T$\mathrm{_{s}}$
where a kink is visible in the data taken with field along $c$. (b)
Pressure dependence of the field cooled data, measured with 1000~Oe
along the $c$ direction. Inset shows extracted values of T$\mathrm{_{c}}$.
(c) Field-dependent ionic magnetic moment, showing clear ferromagnetic
hysteresis loops below T$\mathrm{_{c}}$ - inset shows the orientation
dependence at 2 K.}
\end{figure}

We next explore the magnetic properties of single crystal VI$_{3}$.
The results of measurements are shown in Fig. \ref{fig:MagnetisationPlots}
and additional data given in the Fig 10. Measurements were carried out
with the magnetic field both parallel and perpendicular to the crystallographic
$c$ axis, which is easily identified as perpendicular to the $ab$
crystal planes this two-dimensional material forms. Orientation-dependent
magnetic susceptibility is shown in Fig. \ref{fig:MagnetisationPlots}(a),
measured during warming after cooling to 2~K with the measurement
magnetic field of 100~Oe applied, i.e. field-cooled (FC). Similarly
to the previous result of Wilson et al. \citep{Wilson1987} we find
VI$_{3}$ to be ferromagnetic with T$\mathrm{_{c}}$ of 50.0(1)~K,
defined from the derivative of the susceptibility. 

As shown in Fig. \ref{fig:MagnetisationPlots}(a), there is a strong
anisotropy in the susceptibility - magnetic field applied in-plane
($H\bot c$) leads to magnetization roughly half that found when it
is applied out of plane ($H\parallel c$). We can conclude that VI$_{3}$
has Ising-type spins with the easy axis along $c$. A small kink in
the magnetic susceptibility, as well as a change in slope, in the
paramagnetic state at 79.3~K was observed, corresponding to the structural
transition temperature T$\mathrm{_{s}}$. A structural phase transition
accompanies similar features in the susceptibility of CrI$_{3}$ and
$\alpha$-TiCl$_{3}$ \citep{Tsutsumi1990,McGuire2015}. The kink
was not seen when the field was applied in-plane, but was clearly
present with the field along the easy axis, implying the coupling
of the structural phase transition and the magnetism and a link to
the magnetic anisotropy. 

Zero-field-cooled data are presented in the Fig 10 - a linear Curie-Weiss fit
to these data above T$\mathrm{_{s}}$ allows us to extract an effective
moment $\mu_{eff}=2.08(20)\:\mu_{B}$. Wilson et.al. reported a value
of $2.22\:\mu_{B}$, which agrees with our fit within error and our
ab-initio calculations give a moment of $2\:\mu_{B}$. The extrapolated
Curie Temperature from the fit is 64.5(5)~K, which differs from the
observed T$\mathrm{_{c}}$ as the structural transition at T$\mathrm{_{s}}$
alters the slope of the susceptibility and destabilizes magnetic order.

The effect of applied hydrostatic pressure on the FC susceptibility
in a 1000~Oe field applied along $c$ is shown in Fig \ref{fig:MagnetisationPlots}(b).
Along with the evolution of several bumps in the data most likely
due to domain dynamics, the key result shown in the inset is the behavior
of T$\mathrm{_{c}}$. The effect of pressure on a vdW material such
as VI$_{3}$ will be predominately to continuously decrease inter-layer
spacing, increasing inter-planar exchange and tuning towards a three-dimensional
system. At pressures up to around 7~kbar however, no change in T$\mathrm{_{c}}$
was observed, but at higher pressures it begins to rapidly increase.
We interpret this as evidence of the true two-dimensional nature of
the magnetic order in VI$_{3}$. Bringing the planes closer together
has zero effect on stabilizing the magnetic order to higher temperatures
- inter-planar interactions are unaffected and hence are presumed
to be negligible to begin with. Only above 8~kbar does changing inter-layer
spacing lead to easier formation of magnetic order - at this point
the dimensionality is starting to tune away from 2. The only prior
pressure study, to our knowledge, on the MX$_3$ materials to date
is the work of Yoshida et.al. \citep{Yoshida1997}, which reports
a continuous decrease in the value of T$\mathrm{_{c}}$ with increasing
pressure - the opposite effect to our observations. 

Characteristic ferromagnetic hysteresis loops of the magnetic moment
with applied field along the $c$ axis are shown in Fig. \ref{fig:MagnetisationPlots}(c)
for several temperatures. As temperature is increased,
the coercive field is exponentially decreased (Fig. 10) and above T$\mathrm{_{c}}$
the hysteresis loops close and the system reverts to paramagnetic
linear behavior. At the maximum applied field of 70~kOe (7~T) and
at 2~K the magnetic moment was found to be 1.3(1) $\mu_{B}/\mathrm{f.u.}$,
which is smaller than the expected saturated moment of a V$^{3+}$
ion, $gS=2\mu_{B}/\mathrm{f.u.}$. The moment is not fully saturated
at 70~kOe, but clearly cannot reach this predicted value. The reason
for this disagreement is not currently clear.

The orientation dependence of the magnetization is illustrated in
the inset of Fig. \ref{fig:MagnetisationPlots}(c). The field-in-plane
curves show coercivity of 1.8~kOe, smaller than that of out-of-plane
(9.1~kOe), which again suggests the easy axis to lie along the $c$-axis,
from the Stoner-Wohlfarth model \citep{Stoner1948}. This simple model, it should be noted, while offering valuable insight into the overall behavior strictly applies only to a single-domain crystal - more complex treatments \citep{Ryabchenko2014} may prove more suitable for these data going forwards.  One thing to
note here is its huge coercive field at 2~K - 9.1~kOe, in contrast
to other vdW ferromagnets such as CrBr$_{3}$\citep{Richter2018},
CrI$_{3}$\citep{McGuire2015}, Fe$_{3}$GeTe$_{2}$\citep{Leon-Brito2016}
and Cr$_{2}$Ge$_{2}$Te$_{6}$\citep{Gong2017}. To our knowledge,
all other such ferromagnets are soft, contrary to this hard vdW ferromagnet
VI$_{3}$. Understanding the physical origin of the hard-ferromagnetic
behavior and the comparison of otherwise similar systems VI$_{3}$
and CrI$_{3}$ forms a rich opportunity for low-dimensional magnetism
going forwards and may help in designing future magnetic materials
for specific applications. Interestingly, Chang et al.\citep{Chang2015}
reported in their comparison of V- and Cr-doped Sb$_{2}$Te$_{3}$
that the V-doped samples show much higher coercive fields than the
Cr-doped systems. The Stoner-Wohlfarth model describes that a small
saturated moment, $M_{S}$, and/or a large total magnetic anisotropy
$K$ lead to larger coercive fields. The smaller saturated moment
in V$^{3+}$ driven by the smaller number of d-orbital spin and the
larger anisotropy coming from the partially filled $t_{2g}$ $d$-band
of the V$^{3+}$ ion, unlike the fully filled Cr$^{3+}$ orbitals,
could contribute the large anisotropy, and in turn coercive field. 

\begin{figure}
\centering{}\includegraphics[width=1\columnwidth]{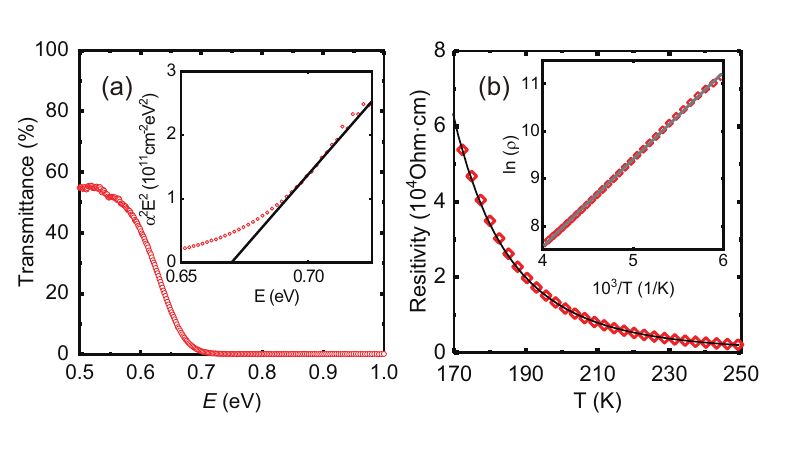}\caption{\label{fig:ResisitivityAndBandGap}(a) Optical transmittance of VI$_{3}$ at room temperature
with inset showing the accompanying Tauc plot and linear fit to extract
the optical band gap. (b) Resistivity, with good agreement to the
Arrhenius insulating temperature dependence shown in the inset. }
\end{figure}

To investigate the electronic properties, we measured the bandgap
by optical transmittance and the temperature dependence of the bulk
resistivity. Fig. \ref{fig:ResisitivityAndBandGap}(a) shows the transmittance at room temperature
as a function of incident photon energy, from which we can extract
the optical bandgap. The dependence of the absorption coefficient
$\alpha$ on incident photon energy is given by the expression $\alpha E\propto(E-E_{g}){}^{m}$
where $E=h\nu$ is the incident photon energy, $E_{g}$ is the optical
band gap and for a direct allowed transition the exponent can be taken
as $m=\nicefrac{1}{2}$. For $E<E_{g}$, the material cannot absorb
photons so the bandgap is found from extrapolating $(\alpha h\nu)^{2}$
vs. $h\nu$ to zero (inset). The obtained direct optical bandgap value
for VI$_{3}$ was 0.67(1)~eV. 

The temperature dependence of the in-plane resistivity, $\rho$, of a bulk single crystal of VI$_{3}$ is plotted
in Fig. \ref{fig:ResisitivityAndBandGap}(b) from room temperature
down to the point where the resistance becomes too high to measure
on our setup. The increasing resistivity as temperature was lowered
shows clear insulating behavior. The resistivity can be fitted well
by an Arrhenius type exponential function: $\rho\propto e^{E_{a}/k_{B}T}$,
where $E_{a}$ is the activation energy and $k_{B}$ Boltzmann's constant.
The inset illustrates the standard $ln(\rho)$ vs. $1/T$ plot - the
straight line proves good agreement to Arrhenius-type thermally activated
transport. The extracted activation energy $E_{a}$ was 0.16~eV,
giving an electrical bandgap of 0.32~eV - significantly smaller than
the measured optical bandgap. The reason for this (common) mismatch is not immediately clear in this
material; a potential explanation is the presence of some form of
impurity band within the gap or denaturing of the surface contact - iodine deficiency or the effect of
exposure to oxygen or moisture may be possible candidates.

Sister compound CrI$_{3}$ is a Mott insulator \citep{McGuire2015}
, with all the rich electronic correlation that implies. Previous results
from ab-initio calculations \citep{He2016} predict VI$_{3}$ to be
metallic, rather than the observed insulating behavior - giving a
hint that VI$_{3}$ is also a Mott-insulating system. Our band structure
calculations (Fig 5) implemeted in the Quantum ESPRESSO package \citep{Gianozzi2009, Dudarev1998, Hamann2013, Schlipf2015, Perdew1996, Monkhorst1976}, which use our updated crystal structure parameters
and include the effects of a Hubbard $U$, suggest a ground state
that is indeed Mott insulating, rather than metallic, with a \textasciitilde{}1~eV
band gap in good agreement with observed data. Reducing the on-site
Coulomb interaction leads to half-metallicity in our calculations. 

We have presented an overview of the basic properties of near-unexplored
van-der-Waals ferromagnet VI$_{3}$. We suggest an updated crystal
structure - crucial for accurate ab-initio calculations - and a transition
into a distorted alternative structure at low temperature. A key result
is a coercive field far higher than in any other vdW ferromagnet system,
setting VI$_{3}$ apart as a hard vdW ferromagnet and raising several
questions worthy of further exploration. Optical and electrical transport
measurements show this material to be an insulator with an optical
band gap of 0.67 eV. In contrast to previously published studies,
our band structure calculations yield an insulating ferromagnetic
ground state when an on-site Coulomb interaction is included, leading
us to believe VI$_{3}$ to be a correlated Mott insulator. This evidence
of Mott physics evokes potential for bandgap tuning and the emergence
of exotic sates due to strong electron correlations in this 2D ferromagnetic
system.

\begin{acknowledgments}
The authors would like to thank Sanghyun Lee, S.E. Dutton, Inho Hwang
and Y. Noda for their generous help and discussions. We would also
like to acknowledge support from Jesus College of the University of
Cambridge, IHT KAZATOMPROM and the CHT Uzbekistan programme. The work
was carried out with financial support from the Ministry of Education
and Science of the Russian Federation in the framework of Increase
Competitiveness Program of NUST MISiS (\textnumero{} \textcyr{\CYRK}2-2017-024).
The work of D.Kh. was supported by the German research project SFB
1238 and by K{\"o}ln University via the German Excellence Initiative.
This work was supported by the van der Waals Materials Research Center
NRF-2017R1A5A1014862 and the Institute for Basic Science of the Republic of Korea (Grant No. IBS-R009-G1).
\end{acknowledgments}

\bibliographystyle{apsrev4-1}

\begin{thebibliography}{1}	
	\bibitem [{ {Ajayan}\ \emph {et~al.}(2016) {Ajayan},
   {Kim},\ and\  {Banerjee}}]{Ajayan2016}%
  \BibitemOpen
  \bibfield  {author} {\bibinfo {author} { {P.}~
  {Ajayan}}, \bibinfo {author} { {P.}~ {Kim}}, \ and\
  \bibinfo {author} { {K.}~ {Banerjee}},\ }\href
  {\doibase 10.1063/pt.3.3297} {\bibfield  {journal} {\bibinfo  {journal}
  {Physics Today}\ }\textbf {\bibinfo {volume} {69}},\ \bibinfo {pages} {38}
  (\bibinfo {year} {2016})}\BibitemShut {NoStop}%
\bibitem [{ {Park}(2016)}]{Park2016}%
  \BibitemOpen
  \bibfield  {author} {\bibinfo {author} { {J.-G.}\ 
  {Park}},\ }\href {\doibase 10.1088/0953-8984/28/30/301001} {\bibfield
  {journal} {\bibinfo  {journal} {J. Phys.: Condens. Matter}\ }\textbf
  {\bibinfo {volume} {28}},\ \bibinfo {pages} {301001} (\bibinfo {year}
  {2016})}\BibitemShut {NoStop}%
\bibitem [{ {Kuo}\ \emph {et~al.}(2016) {Kuo},
   {Neumann},  {Balamurugan},  {Park},
   {Kang},  {Shiu},  {Kang},
   {Hong},  {Han},  {Noh},\ and\
   {Park}}]{Kuo2016}%
  \BibitemOpen
  \bibfield  {author} {\bibinfo {author} { {C.}~
  {Kuo}}, \bibinfo {author} { {M.}~ {Neumann}},
  \bibinfo {author} { {K.}~ {Balamurugan}}, \bibinfo
  {author} { {H.}~ {Park}}, \bibinfo {author}
  { {S.}~ {Kang}}, \bibinfo {author} {
  {H.}~ {Shiu}}, \bibinfo {author} { {J.}~
  {Kang}}, \bibinfo {author} { {B.}~ {Hong}}, \bibinfo
  {author} { {M.}~ {Han}}, \bibinfo {author}
  { {T.}~ {Noh}}, \ and\ \bibinfo {author}
  { {J.-G.}\  {Park}},\ }\href {\doibase
  10.1038/srep20904} {\bibfield  {journal} {\bibinfo  {journal} {Scientific
  Reports}\ }\textbf {\bibinfo {volume} {6}},\ \bibinfo {pages} {20904}
  (\bibinfo {year} {2016})}\BibitemShut {NoStop}%
\bibitem [{ {Zhou}\ \emph {et~al.}(2016) {Zhou},
   {Lu},  {Zu},\ and\ 
  {Gao}}]{Zhou2016}%
  \BibitemOpen
  \bibfield  {author} {\bibinfo {author} { {Y.}~
  {Zhou}}, \bibinfo {author} { {H.}~ {Lu}}, \bibinfo
  {author} { {X.}~ {Zu}}, \ and\ \bibinfo {author}
  { {F.}~ {Gao}},\ }\href {\doibase 10.1038/srep19407}
  {\bibfield  {journal} {\bibinfo  {journal} {Scientific Reports}\ }\textbf
  {\bibinfo {volume} {6}} (\bibinfo {year} {2016}),\
  10.1038/srep19407}\BibitemShut {NoStop}%
\bibitem [{ {Samarth}(2017)}]{Samarth2017}%
  \BibitemOpen
  \bibfield  {author} {\bibinfo {author} { {N.}~
  {Samarth}},\ }\href {\doibase 10.1038/546216a} {\bibfield  {journal}
  {\bibinfo  {journal} {Nature}\ }\textbf {\bibinfo {volume} {546}},\ \bibinfo
  {pages} {216} (\bibinfo {year} {2017})}\BibitemShut {NoStop}%
\bibitem [{ {Burch}\ \emph {et~al.}(2018) {Burch},
   {Mandrus},\ and\  {Park}}]{Burch2018}%
  \BibitemOpen
  \bibfield  {author} {\bibinfo {author} { {K.}~
  {Burch}}, \bibinfo {author} { {D.}~ {Mandrus}}, \
  and\ \bibinfo {author} { {J.-G.}\  {Park}},\ }\href
  {\doibase 10.1038/s41586-018-0631-z} {\bibfield  {journal} {\bibinfo
  {journal} {Nature}\ }\textbf {\bibinfo {volume} {563}},\ \bibinfo {pages}
  {47} (\bibinfo {year} {2018})}\BibitemShut {NoStop}%
\bibitem [{ {McGuire}\ \emph {et~al.}(2015)
  {McGuire},  {Dixit},  {Cooper},\ and\ 
  {Sales}}]{McGuire2015}%
  \BibitemOpen
  \bibfield  {author} {\bibinfo {author} { {M.}~
  {McGuire}}, \bibinfo {author} { {H.}~ {Dixit}},
  \bibinfo {author} { {V.}~ {Cooper}}, \ and\ \bibinfo
  {author} { {B.}~ {Sales}},\ }\href {\doibase
  10.1021/cm504242t} {\bibfield  {journal} {\bibinfo  {journal} {Chemistry of
  Materials}\ }\textbf {\bibinfo {volume} {27}},\ \bibinfo {pages} {612}
  (\bibinfo {year} {2015})}\BibitemShut {NoStop}%
\bibitem [{ {Zhang}\ \emph {et~al.}(2015) {Zhang},
   {Qu},  {Zhu},\ and\ 
  {Lam}}]{Zhang2015}%
  \BibitemOpen
  \bibfield  {author} {\bibinfo {author} { {W.-B.}\ 
  {Zhang}}, \bibinfo {author} { {Q.}~ {Qu}}, \bibinfo
  {author} { {P.}~ {Zhu}}, \ and\ \bibinfo {author}
  { {C.-H.}\  {Lam}},\ }\href {\doibase
  10.1039/c5tc02840j} {\bibfield  {journal} {\bibinfo  {journal} {Journal of
  Materials Chemistry C}\ }\textbf {\bibinfo {volume} {3}},\ \bibinfo {pages}
  {12457} (\bibinfo {year} {2015})}\BibitemShut {NoStop}%
\bibitem [{ {Wang}\ \emph {et~al.}(2016) {Wang},
   {Fan},  {Zhu},\ and\ 
  {Wu}}]{Wang2016}%
  \BibitemOpen
  \bibfield  {author} {\bibinfo {author} { {H.}~
  {Wang}}, \bibinfo {author} { {F.}~ {Fan}}, \bibinfo
  {author} { {S.}~ {Zhu}}, \ and\ \bibinfo {author}
  { {H.}~ {Wu}},\ }\href {\doibase
  10.1209/0295-5075/114/47001} {\bibfield  {journal} {\bibinfo  {journal}
  {{EPL} (Europhysics Letters)}\ }\textbf {\bibinfo {volume} {114}},\ \bibinfo
  {pages} {47001} (\bibinfo {year} {2016})}\BibitemShut {NoStop}%
\bibitem [{ {Huang}\ \emph {et~al.}(2017) {Huang},
   {Clark},  {Navarro-Moratalla}, 
  {Klein},  {Cheng},  {Seyler}, 
  {Zhong},  {Schmidgall},  {McGuire}, 
  {Cobden},  {Yao},  {Xiao}, 
  {Jarillo-Herrero},\ and\  {Xu}}]{Huang2017}%
  \BibitemOpen
  \bibfield  {author} {\bibinfo {author} { {B.}~
  {Huang}}, \bibinfo {author} { {G.}~ {Clark}},
  \bibinfo {author} { {E.}~ {Navarro-Moratalla}},
  \bibinfo {author} { {D.}~ {Klein}}, \bibinfo
  {author} { {R.}~ {Cheng}}, \bibinfo {author}
  { {K.}~ {Seyler}}, \bibinfo {author} {
  {D.}~ {Zhong}}, \bibinfo {author} {
  {E.}~ {Schmidgall}}, \bibinfo {author} { {M.~A.}\
   {McGuire}}, \bibinfo {author} { {D.~H.}\
   {Cobden}}, \bibinfo {author} { {W.}~
  {Yao}}, \bibinfo {author} { {D.}~ {Xiao}}, \bibinfo
  {author} { {P.}~ {Jarillo-Herrero}}, \ and\ \bibinfo
  {author} { {X.}~ {Xu}},\ }\href {\doibase
  10.1038/nature22391} {\bibfield  {journal} {\bibinfo  {journal} {Nature}\
  }\textbf {\bibinfo {volume} {546}},\ \bibinfo {pages} {270} (\bibinfo {year}
  {2017})}\BibitemShut {NoStop}%
\bibitem [{ {Lado}\ and\ 
  {Fern{\'{a}}ndez-Rossier}(2017)}]{Lado2017}%
  \BibitemOpen
  \bibfield  {author} {\bibinfo {author} { {J.~L.}\ 
  {Lado}}\ and\ \bibinfo {author} { {J.}~
  {Fern{\'{a}}ndez-Rossier}},\ }\href {\doibase 10.1088/2053-1583/aa75ed}
  {\bibfield  {journal} {\bibinfo  {journal} {2D Materials}\ }\textbf {\bibinfo
  {volume} {4}},\ \bibinfo {pages} {035002} (\bibinfo {year}
  {2017})}\BibitemShut {NoStop}%
\bibitem [{ {Klein}\ \emph {et~al.}(2018) {Klein},
   {MacNeill},  {Lado},  {Soriano},
   {Navarro-Moratalla},  {Watanabe}, 
  {Taniguchi},  {Manni},  {Canfield}, 
  {Fern{\'{a}}ndez-Rossier},\ and\ 
  {Jarillo-Herrero}}]{Klein2018}%
  \BibitemOpen
  \bibfield  {author} {\bibinfo {author} { {D.~R.}\ 
  {Klein}}, \bibinfo {author} { {D.}~ {MacNeill}},
  \bibinfo {author} { {J.~L.}\  {Lado}}, \bibinfo
  {author} { {D.}~ {Soriano}}, \bibinfo {author}
  { {E.}~ {Navarro-Moratalla}}, \bibinfo {author}
  { {K.}~ {Watanabe}}, \bibinfo {author}
  { {T.}~ {Taniguchi}}, \bibinfo {author}
  { {S.}~ {Manni}}, \bibinfo {author} {
  {P.}~ {Canfield}}, \bibinfo {author} {
  {J.}~ {Fern{\'{a}}ndez-Rossier}}, \ and\ \bibinfo {author}
  { {P.}~ {Jarillo-Herrero}},\ }\href {\doibase
  10.1126/science.aar3617} {\bibfield  {journal} {\bibinfo  {journal}
  {Science}\ }\textbf {\bibinfo {volume} {360}},\ \bibinfo {pages} {1218}
  (\bibinfo {year} {2018})}\BibitemShut {NoStop}%
\bibitem [{ {Juza}\ \emph {et~al.}(1969) {Juza},
   {Giegling},\ and\  {Sch\"afer}}]{Juza1969}%
  \BibitemOpen
  \bibfield  {author} {\bibinfo {author} { {D.}~
  {Juza}}, \bibinfo {author} { {D.}~ {Giegling}}, \
  and\ \bibinfo {author} { {H.}~{Sch\"afer}},\
  }\href {\doibase 10.1002/zaac.19693660303} {\bibfield  {journal} {\bibinfo
  {journal} {Zeitschrift f\"ur anorganische und allgemeine Chemie}\ }\textbf
  {\bibinfo {volume} {366}},\ \bibinfo {pages} {121} (\bibinfo {year}
  {1969})}\BibitemShut {NoStop}%
\bibitem [{ {Dillon}\ and\ 
  {Olson}(1965)}]{Dillon1965}%
  \BibitemOpen
  \bibfield  {author} {\bibinfo {author} { {J.}~
  {Dillon}}\ and\ \bibinfo {author} { {C.}~ {Olson}},\
  }\href {\doibase 10.1063/1.1714194} {\bibfield  {journal} {\bibinfo
  {journal} {Journal of Applied Physics}\ }\textbf {\bibinfo {volume} {36}},\
  \bibinfo {pages} {1259} (\bibinfo {year} {1965})}\BibitemShut {NoStop}%
\bibitem [{ {McGuire}(2017)}]{McGuire2017}%
  \BibitemOpen
  \bibfield  {author} {\bibinfo {author} { {M.}~
  {McGuire}},\ }\href {\doibase 10.3390/cryst7050121} {\bibfield  {journal}
  {\bibinfo  {journal} {Crystals}\ }\textbf {\bibinfo {volume} {7}},\ \bibinfo
  {pages} {121} (\bibinfo {year} {2017})}\BibitemShut {NoStop}%
\bibitem [{ {Trotter}\ and\ 
  {Zobel}(1966)}]{Trotter1966}%
  \BibitemOpen
  \bibfield  {author} {\bibinfo {author} { {J.}~
  {Trotter}}\ and\ \bibinfo {author} { {T.}~
  {Zobel}},\ }\href {\doibase 10.1524/zkri.1966.123.16.67} {\bibfield
  {journal} {\bibinfo  {journal} Zeitschrift f{\"u}r {K}ristallographie -
  {Crystalline Materials}\ }\textbf {\bibinfo {volume} {123}} (\bibinfo {year}
  {1966}),\ 10.1524/zkri.1966.123.16.67}\BibitemShut {NoStop}%
\bibitem [{ {Handy}(1950)}]{Handy1950}%
  \BibitemOpen
  \bibfield  {author} {\bibinfo {author} { {L.L.}~{Handy}} and \bibinfo {author} { {N.W.}~{Gregory}}} {\bibfield  {journal}
  {\bibinfo  {journal} {Journal of the American Chemical Society}\ }\textbf {\bibinfo {volume} {72}},\ \bibinfo
  {pages} {5049-5051} (\bibinfo {year} {1950})}
\BibitemShut {NoStop}%
\bibitem [{ {Wilson}\ \emph {et~al.}(1987) {Wilson},
   {Maule},  {Strange},\ and\ 
  {Tothill}}]{Wilson1987}%
  \BibitemOpen
  \bibfield  {author} {\bibinfo {author} { {J.}~
  {Wilson}}, \bibinfo {author} { {C.}~ {Maule}},
  \bibinfo {author} { {P.}~ {Strange}}, \ and\
  \bibinfo {author} { {J.}~ {Tothill}},\ }\href
  {\doibase 10.1088/0022-3719/20/26/017} {\bibfield  {journal} {\bibinfo
  {journal} {Journal of Physics C: Solid State Physics}\ }\textbf {\bibinfo
  {volume} {20}},\ \bibinfo {pages} {4159} (\bibinfo {year}
  {1987})}\BibitemShut {NoStop}%
\bibitem [{ {He}\ \emph {et~al.}(2016) {He},
   {Ma},  {Lyu},\ and\ 
  {Nachtigall}}]{He2016}%
  \BibitemOpen
  \bibfield  {author} {\bibinfo {author} { {J.}~
  {He}}, \bibinfo {author} { {S.}~ {Ma}}, \bibinfo
  {author} { {P.}~ {Lyu}}, \ and\ \bibinfo {author}
  { {P.}~ {Nachtigall}},\ }\href {\doibase
  10.1039/c6tc00409a} {\bibfield  {journal} {\bibinfo  {journal} {Journal of
  Materials Chemistry C}\ }\textbf {\bibinfo {volume} {4}},\ \bibinfo {pages}
  {2518} (\bibinfo {year} {2016})}\BibitemShut {NoStop}%
\bibitem [{ {{Zheng}}\ \emph {et~al.}(2017)
  {{Zheng}},  {{Zhao}},  {{Liu}}, 
  {{Li}},  {{Zhou}},  {{Zhang}},\ and\ 
  {{Zhang}}}]{Zheng2017}%
  \BibitemOpen
  \bibfield  {author} {\bibinfo {author} { {F.}~
  {{Zheng}}}, \bibinfo {author} { {J.}~ {{Zhao}}},
  \bibinfo {author} { {Z.}~ {{Liu}}}, \bibinfo
  {author} { {M.}~ {{Li}}}, \bibinfo {author}
  { {M.}~ {{Zhou}}}, \bibinfo {author} {
  {S.}~ {{Zhang}}}, \ and\ \bibinfo {author} {
  {P.}~ {{Zhang}}},\ }\href {https://arxiv.org/abs/1709.05472}
  {\bibfield  {journal} {\bibinfo  {journal} {ArXiv e-prints}\ } {\bibinfo  {volume} {1709.05472}\ } (\bibinfo
  {year} {2017})},\  \BibitemShut {NoStop}%
\bibitem [{ {Berry}\ \emph {et~al.}(1969) {Berry},
   {Smardzewski},\ and\  {McCarley}}]{Berry1969}%
  \BibitemOpen
  \bibfield  {author} {\bibinfo {author} { {K.}~
  {Berry}}, \bibinfo {author} { {R.}~ {Smardzewski}},
  \ and\ \bibinfo {author} { {R.}~ {McCarley}},\
  }\href {\doibase 10.1021/ic50079a034} {\bibfield  {journal} {\bibinfo
  {journal} {Inorganic Chemistry}\ }\textbf {\bibinfo {volume} {8}},\ \bibinfo
  {pages} {1994} (\bibinfo {year} {1969})}\BibitemShut {NoStop}%
\bibitem [{ {Rodr\'iguez-Carvajal}(1993)}]{Rodriguez1993}%
  \BibitemOpen
  \bibfield  {author} {\bibinfo {author} { {J.}~
  {Rodr\'iguez-Carvajal}},\ } {\bibfield  {journal}
  {\bibinfo  {journal} {Physica B: Condensed Matter}\ }\textbf {\bibinfo {volume} {192}},\ \bibinfo
  {pages} {55-69} (\bibinfo {year} {1993})}
\BibitemShut {NoStop}%
\bibitem [{ {Toby}(2013)}]{Toby2013}%
  \BibitemOpen
  \bibfield  {author} {\bibinfo {author} { {B.H.}~{Toby}} and \bibinfo {author} { {R.B.}~{Von Dreele}}} {\bibfield  {journal}
  {\bibinfo  {journal} {Journal of Applied Crystallography}\ }\textbf {\bibinfo {volume} {46}},\ \bibinfo
  {pages} {544-549} (\bibinfo {year} {2013})}
\BibitemShut {NoStop}%
\bibitem [{ {Momma}(2011)}]{Momma2011}%
  \BibitemOpen
  \bibfield  {author} {\bibinfo {author} { {K.}~{Momma}} and \bibinfo {author} { {F.}~{Izumi}}} {\bibfield  {journal}
  {\bibinfo  {journal} {Journal of Applied Crystallography}\ }\textbf {\bibinfo {volume} {44}},\ \bibinfo
  {pages} {1272-1276} (\bibinfo {year} {2011})}
\BibitemShut {NoStop}%
\bibitem [{ {Natta}\ \emph {et~al.}(1961) {Natta},
   {Corradini},\ and\  {Allegra}}]{Natta1961}%
  \BibitemOpen
  \bibfield  {author} {\bibinfo {author} { {G.}~
  {Natta}}, \bibinfo {author} { {P.}~ {Corradini}}, \
  and\ \bibinfo {author} { {G.}~ {Allegra}},\ }\href
  {\doibase 10.1002/pol.1961.1205115602} {\bibfield  {journal} {\bibinfo
  {journal} {Journal of Polymer Science}\ }\textbf {\bibinfo {volume} {51}},\
  \bibinfo {pages} {399} (\bibinfo {year} {1961})}\BibitemShut {NoStop}%
\bibitem [{ {Hillebrecht}\ \emph {et~al.}(1997)
  {Hillebrecht},  {Schmidt},  {Rotter}, 
  {Thiele},  {Z\"onnchen},  {Bengel}, 
  {Cantow},  {Magonov},\ and\ 
  {Whangbo}}]{Hillebrecht1997}%
  \BibitemOpen
  \bibfield  {author} {\bibinfo {author} { {H.}~
  {Hillebrecht}}, \bibinfo {author} { {P.}~
  {Schmidt}}, \bibinfo {author} { {H.}~ {Rotter}},
  \bibinfo {author} { {G.}~ {Thiele}}, \bibinfo
  {author} { {P.}~ {Z\"onnchen}}, \bibinfo {author}
  { {H.}~ {Bengel}}, \bibinfo {author} {
  {H.-J.}\  {Cantow}}, \bibinfo {author} {
  {S.}~ {Magonov}}, \ and\ \bibinfo {author} {
  {M.-H.}\  {Whangbo}},\ }\href {\doibase
  10.1016/s0925-8388(96)02465-6} {\bibfield  {journal} {\bibinfo  {journal}
  {Journal of Alloys and Compounds}\ }\textbf {\bibinfo {volume} {246}},\
  \bibinfo {pages} {70} (\bibinfo {year} {1997})}\BibitemShut {NoStop}%
\bibitem [{ {Novoselov}\ \emph {et~al.}(2004)
  {Novoselov},  {Geim},  {Morozov}, 
  {Jiang},  {Zhang},  {Dubonos}, 
  {Grigrieva},\ and\  {Firsov}}]{Novoselov2004}%
  \BibitemOpen
  \bibfield  {author} {\bibinfo {author} { {K.~S.}\ 
  {Novoselov}}, \bibinfo {author} { {A.}~ {Geim}},
  \bibinfo {author} { {S.~V.}\  {Morozov}}, \bibinfo
  {author} { {D.}~ {Jiang}}, \bibinfo {author}
  { {Y.}~ {Zhang}}, \bibinfo {author} {
  {S.~V.}\  {Dubonos}}, \bibinfo {author} { {I.~V.}\
   {Grigrieva}}, \ and\ \bibinfo {author} { {A.~A.}\
   {Firsov}},\ }\href {\doibase 10.1126/science.1102896} {\bibfield
   {journal} {\bibinfo  {journal} {Science}\ }\textbf {\bibinfo {volume}
  {306}},\ \bibinfo {pages} {666} (\bibinfo {year} {2004})}\BibitemShut
  {NoStop}%
\bibitem [{ {Tsutsumi}\ \emph {et~al.}(1990)
  {Tsutsumi},  {Okamoto},  {Hama},\ and\
   {Ishihara}}]{Tsutsumi1990}%
  \BibitemOpen
  \bibfield  {author} {\bibinfo {author} { {K.}~
  {Tsutsumi}}, \bibinfo {author} { {H.}~ {Okamoto}},
  \bibinfo {author} { {C.}~ {Hama}}, \ and\ \bibinfo
  {author} { {Y.}~ {Ishihara}},\ }\href {\doibase
  10.1016/s0304-8853(10)80063-0} {\bibfield  {journal} {\bibinfo  {journal}
  {Journal of Magnetism and Magnetic Materials}\ }\textbf {\bibinfo {volume}
  {90-91}},\ \bibinfo {pages} {181} (\bibinfo {year} {1990})}\BibitemShut
  {NoStop}%
\bibitem [{ {Yoshida}\ \emph {et~al.}(1997)
  {Yoshida},  {Chiba},  {Kaneko}, 
  {Fujimori},\ and\  {Abe}}]{Yoshida1997}%
  \BibitemOpen
  \bibfield  {author} {\bibinfo {author} { {H.}~
  {Yoshida}}, \bibinfo {author} { {J.}~ {Chiba}},
  \bibinfo {author} { {T.}~ {Kaneko}}, \bibinfo
  {author} { {Y.}~ {Fujimori}}, \ and\ \bibinfo
  {author} { {S.}~ {Abe}},\ }\href {\doibase
  10.1016/s0921-4526(97)00207-x} {\bibfield  {journal} {\bibinfo  {journal}
  {Physica B: Condensed Matter}\ }\textbf {\bibinfo {volume} {237-238}},\
  \bibinfo {pages} {525} (\bibinfo {year} {1997})}\BibitemShut {NoStop}%
\bibitem [{ {Stoner}\ and\ 
  {Wohlfarth}(1948)}]{Stoner1948}%
  \BibitemOpen
  \bibfield  {author} {\bibinfo {author} { {E.}~
  {Stoner}}\ and\ \bibinfo {author} { {E.}~
  {Wohlfarth}},\ }\href {\doibase 10.1098/rsta.1948.0007} {\bibfield  {journal}
  {\bibinfo  {journal} {Philosophical Transactions of the Royal Society A:
  Mathematical, Physical and Engineering Sciences}\ }\textbf {\bibinfo {volume}
  {240}},\ \bibinfo {pages} {599} (\bibinfo {year} {1948})}\BibitemShut
  {NoStop}%
\bibitem [{ {Richter}\ \emph {et~al.}(2018)
  {Richter},  {Weber},  {Martin}, 
  {Singh},  {Schwingenschl\:{o}gl},  {Lotsch},\ and\
   {Kl?ui}}]{Richter2018}%
  \BibitemOpen
  \bibfield  {author} {\bibinfo {author} { {N.}~
  {Richter}}, \bibinfo {author} { {D.}~ {Weber}},
  \bibinfo {author} { {F.}~ {Martin}}, \bibinfo
  {author} { {N.}~ {Singh}}, \bibinfo {author}
   {U.}~{S}chwingenschl\"{o}gl, \bibinfo {author}
  { {B.~V.}\  {Lotsch}}, \ and\ \bibinfo {author}
   {M.}~{K}l\"{a}ui,\ }\href {\doibase
  10.1103/physrevmaterials.2.024004} {\bibfield  {journal} {\bibinfo  {journal}
  {Physical Review Materials}\ }\textbf {\bibinfo {volume} {2}} (\bibinfo
  {year} {2018}),\ 10.1103/physrevmaterials.2.024004}\BibitemShut {NoStop}%
\bibitem [{ {Le{\'{o}}n-Brito}\ \emph {et~al.}(2016)
  {Le{\'{o}}n-Brito},  {Bauer},  {Ronning},
   {Thompson},\ and\  {Movshovich}}]{Leon-Brito2016}%
  \BibitemOpen
  \bibfield  {author} {\bibinfo {author} { {N.}~
  {Le{\'{o}}n-Brito}}, \bibinfo {author} { {E.~D.}\ 
  {Bauer}}, \bibinfo {author} { {F.}~ {Ronning}},
  \bibinfo {author} { {J.~D.}\  {Thompson}}, \ and\
  \bibinfo {author} { {R.}~ {Movshovich}},\ }\href
  {\doibase 10.1063/1.4961592} {\bibfield  {journal} {\bibinfo  {journal}
  {Journal of Applied Physics}\ }\textbf {\bibinfo {volume} {120}},\ \bibinfo
  {pages} {083903} (\bibinfo {year} {2016})}\BibitemShut {NoStop}%
\bibitem [{ {Gong}\ \emph {et~al.}(2017) {Gong},
   {Li},  {Li},  {Ji}, 
  {Stern},  {Xia},  {Cao},  {Bao},
   {Wang},  {Wang},  {Qiu},
   {Cava},  {Louie},  {Xia},\ and\
   {Zhang}}]{Gong2017}%
  \BibitemOpen
  \bibfield  {author} {\bibinfo {author} { {C.}~
  {Gong}}, \bibinfo {author} { {L.}~ {Li}}, \bibinfo
  {author} { {Z.}~ {Li}}, \bibinfo {author}
  { {H.}~ {Ji}}, \bibinfo {author} {
  {A.}~ {Stern}}, \bibinfo {author} {
  {Y.}~ {Xia}}, \bibinfo {author} { {T.}~
  {Cao}}, \bibinfo {author} { {W.}~ {Bao}}, \bibinfo
  {author} { {C.}~ {Wang}}, \bibinfo {author}
  { {Y.}~ {Wang}}, \bibinfo {author} {
  {Z.}~ {Qiu}}, \bibinfo {author} { {R.}~
  {Cava}}, \bibinfo {author} { {S.}~ {Louie}},
  \bibinfo {author} { {J.}~ {Xia}}, \ and\ \bibinfo
  {author} { {X.}~ {Zhang}},\ }\href {\doibase
  10.1038/nature22060} {\bibfield  {journal} {\bibinfo  {journal} {Nature}\
  }\textbf {\bibinfo {volume} {546}},\ \bibinfo {pages} {265} (\bibinfo {year}
  {2017})}\BibitemShut {NoStop}%
\bibitem [{ {Chang}\ \emph {et~al.}(2015) {Chang},
   {Zhao},  {Kim},  {Zhang},
   {Assaf},  {Heiman},  {Zhang},
   {Liu},  {Chan},\ and\ 
  {Moodera}}]{Chang2015}%
  \BibitemOpen
  \bibfield  {author} {\bibinfo {author} { {C.-Z.}\ 
  {Chang}}, \bibinfo {author} { {W.}~ {Zhao}},
  \bibinfo {author} { {D.~Y.}\  {Kim}}, \bibinfo
  {author} { {H.}~ {Zhang}}, \bibinfo {author}
  { {B.~A.}\  {Assaf}}, \bibinfo {author}
  { {D.}~ {Heiman}}, \bibinfo {author} {
  {S.-C.}\  {Zhang}}, \bibinfo {author} {
  {C.}~ {Liu}}, \bibinfo {author} { {M.~H.~W.}\
   {Chan}}, \ and\ \bibinfo {author} { {J.~S.}\
   {Moodera}},\ }\href {\doibase 10.1038/nmat4204} {\bibfield
  {journal} {\bibinfo  {journal} {Nature Materials}\ }\textbf {\bibinfo
  {volume} {14}},\ \bibinfo {pages} {473} (\bibinfo {year} {2015})}\BibitemShut
  {NoStop}%
	\bibitem [{ {Ryabchenko}\ \emph {et~al.}(2014) {Ryabchenko}, and {Moodera}}]{Ryabchenko2014}%
  \BibitemOpen
  \bibfield  {author} {\bibinfo {author} { {S.M.}\ 
  {Ryabchenko}}, \bibinfo {author} { {V.M.}~ {Kalita} }}{\bibfield
  {journal} {\bibinfo  {journal} {Journal of Experimental and Theoretical Physics}\ }\textbf {\bibinfo
  {volume} {118}},\ \bibinfo {pages} {284-296} (\bibinfo {year} {2014})}\BibitemShut{NoStop}%
	\bibitem [{ {Gianozzi}(2009)}]{Gianozzi2009}%
  \BibitemOpen
  \bibfield  {author} {\bibinfo {author} {P. Giannozzi}, \bibinfo {author} {S. Baroni}, \bibinfo {author} {N. Bonini}, \bibinfo {author} {M. Calandra}, \bibinfo {author} {R. Car}, \bibinfo {author} {C. Cavazzoni}, \bibinfo {author} {D. Ceresoli}, \bibinfo {author} {G.L. Chiarotti}, \bibinfo {author} {M. Cococcioni}, \bibinfo {author} {I. Dabo}, \bibinfo {author} {A. Dal Corso}, \bibinfo {author} {S. de Gironcoli}, \bibinfo {author} {S. Fabris}, \bibinfo {author} {G. Fratesi}, \bibinfo {author} {R. Gebauer}, \bibinfo {author} {U. Gerstmann}, \bibinfo {author} {C. Gougoussis}, \bibinfo {author} {A. Kokalj}, \bibinfo {author} {M. Lazzeri}, \bibinfo {author} {L. Martin-Samos}, \bibinfo {author} {N. Marzari}, \bibinfo {author} {F. Mauri}, \bibinfo {author} {R. Mazzarello}, \bibinfo {author} {S. Paolini}, \bibinfo {author} {A. Pasquarello}, \bibinfo {author} {L. Paulatto}, \bibinfo {author} {C. Sbraccia}, \bibinfo {author} {S. Scandolo}, \bibinfo {author} {G. Sclauzero}, \bibinfo {author} {A.P. Seitsonen}, \bibinfo {author} {A. Smogunov}, \bibinfo {author} {P. Umari} and \bibinfo {author} {R.M. Wentzcovitch }} {\bibfield  {journal}
  {\bibinfo  {journal} {J. Phys. Cond. Mat.}\ }\textbf {\bibinfo {volume} {21}},\ \bibinfo
  {pages} {395502} (\bibinfo {year} {2009})}
\BibitemShut {NoStop}%
\bibitem [{ {Dudarev}(1998)}]{Dudarev1998}%
  \BibitemOpen
  \bibfield  {author} {\bibinfo {author} { {S.L.}~{Dudarev}}, \bibinfo {author} { {G.A.}~{Botton}}, \bibinfo {author} { {S.Y.}~{Savrasov}}, \bibinfo {author} { {C.J.}~{Humphreys}} and \bibinfo {author} { {A.P.}~{Sutton}}} {\bibfield  {journal}
  {\bibinfo  {journal} {Phys. Rev. B}\ }\textbf {\bibinfo {volume} {57}},\ \bibinfo
  {pages} {1505-1509} (\bibinfo {year} {1998})}
\BibitemShut {NoStop}%
\bibitem [{ {Hamann}(2013)}]{Hamann2013}%
  \BibitemOpen
  \bibfield  {author} {\bibinfo {author} { {D.R.}~{Hamann}}} {\bibfield  {journal}
  {\bibinfo  {journal} {Phys. Rev. B}\ }\textbf {\bibinfo {volume} {88}} (\bibinfo {year} {2013})}
\BibitemShut {NoStop}%
\bibitem [{ {Schlipf}(2015)}]{Schlipf2015}%
  \BibitemOpen
  \bibfield  {author} {\bibinfo {author} { {M.}~{Schlipf}} and \bibinfo {author} { {F.}~{Gygi}}} {\bibfield  {journal}
  {\bibinfo  {journal} {Computer Physics Comms.}\ }\textbf {\bibinfo {volume} {196}},\ \bibinfo
  {pages} {1272-1276} (\bibinfo {year} {2015})}
\BibitemShut {NoStop}%
\bibitem [{ {Perdew}(1966)}]{Perdew1996}%
  \BibitemOpen
  \bibfield  {author} {\bibinfo {author} { {J.P.}~{Perdew}}, \bibinfo {author} { {K.}~{Burke}} and \bibinfo {author} { {M.}~{Ernzerhof}}} {\bibfield  {journal}
  {\bibinfo  {journal} {Phys. Rev. Lett.}\ }\textbf {\bibinfo {volume} {77}},\ \bibinfo
  {pages} {3865-3868} (\bibinfo {year} {1996})}
\BibitemShut {NoStop}%
\bibitem [{ {Monkhorst}(1976)}]{Monkhorst1976}%
  \BibitemOpen
  \bibfield  {author} {\bibinfo {author} { {H.J.}~{Monkhorst}} and \bibinfo {author} { {J.D.}~{Pack}}} {\bibfield  {journal}
  {\bibinfo  {journal} {Phys. Rev. B}\ }\textbf {\bibinfo {volume} {13}},\ \bibinfo
  {pages} {5188-5192} (\bibinfo {year} {1976})}
\BibitemShut {NoStop}%
\end{thebibliography}

\subsection*{Methods}

Single crystals of VI$_{3}$ were grown via a self-flux chemical vapor
transport method in quartz tubes evacuated to high vacuum, following
the method of Juza et.al.\citep{Juza1969}. Vanadium powder (99.9\%,
Sigma Aldrich) and crystalline iodine (99.99\%, Alfa Aesar) were combined
in the stoichiometric ratio with additional 5\% Iodine within an Ar
atmosphere glove box, and their containing quartz tube then evacuated
to \textasciitilde{}$10^{-2}$ Torr and sealed. The tubes were placed
in a two-zone furnace and heated to 400 C / 320 C over 6 hours. The
furnace was held at these temperatures for 7 days, then cooled to
room temperature over 12 hours. The crystals grown formed shiny black
flakes or platelets, allowing the crystal planes to be easily identified.
The maximum dimension of single crystals was 1 cm{*}1 cm{*}100 um.
Powder samples were grown via solid state reaction in a 5~Torr Ar
atmosphere. The resulting powder was ground and then annealed at 300 C
for 2 days, and then another annealing was made at 400 C for 2 days
after further grinding. At each case, after additional 2 days in Ar
atmosphere for the removal of excessive Iodine in the sample, the
sample stoichiometry and quality were verified with energy dispersive
x-ray (EDX) spectroscopy and powder x-ray diffraction (XRD). 

Temperature dependent powder XRD measurements were carried out using
a Bruker D8 Discovery system with Cu target K$\alpha1$ and K$\alpha2$
wavelengths. Diffraction data were analyzed in the FULLPROF \citep{Rodriguez1993}
and GSAS-II \citep{Toby2013} software suites and visualized in VESTA
\citep{Momma2011}. It was necessary to take into account the effect
of preferred orientation through the (00l) direction in all refinements. 

Resistivity measurements were carried out using a Keithley 6430 source
meter and a closed cycle cryostat system developed in-house. These
were on bulk single crystals with current applied and voltage measured
in the $ab$ crystal plane in a 4-wire geometry. Crystal dimensions
were obtained by inspection under microscope with scale calibrants.
Optical transmittance measurements were done by using an Agilent Cary
5000 spectrophotometer. As with all but the most carefully prepared
CrI$_{3}$ or other halide samples \citep{Handy1950,McGuire2015},
VI$_{3}$ was found to be extremely sensitive to atmospheric moisture.
All sample preparation was therefore carried out in an argon atmosphere.
All measurements of magnetic properties were performed using a Quantum
Design MPMS3 EverCool SQUID magnetometer. Single crystals were characterized
in ranges of temperature (2 to 300~K) and magnetic field (-7 to 7~T)
with the vibrating sample magnetometer option. A high-pressure environment
was achieved within the MPMS by employing a piston-cylinder clamp
cell developed by CamCool Research Ltd, UK. This cell is optimized
for the dimensions of the MPMS and its geometry and gold-plated beryllium-copper
construction are designed for a minimal magnetic background. Daphne
oil 7373 was used as the pressure transmitting medium and the superconducting
transition of lead as a pressure calibrant. The DC data collection
option was used for these measurements, and the background signal
of an empty pressure cell subtracted from the raw data before fitting
the dipole signal. 

The electronic structure of bulk VI$_{3}$ was calculated within density
functional theory (DFT) using the Quantum ESPRESSO package \citep{Gianozzi2009}.
The atomic coordinates of the high-temperature structure (Table \ref{tab:refinementParamsTable})
were used for the calculations. To consider the strong correlation
effects of localized 3d electrons of V atoms, a simplified version
of the DFT+U method suggested by Dudarev et al. \citep{Dudarev1998}
was employed. The effective Hubbard interaction parameter $U_{\mathrm{eff}}$
of V atoms was set to 3.7~eV, consistent with a recent calculation
result based on a linear response approach \citep{He2016}. Interactions
between ion cores and electrons were represented by optimized norm-conserving
pseudopotentials \citep{Hamann2013,Schlipf2015}. The generalized
gradient approximation of Perdew, Burke and Ernzerhof \citep{Perdew1996}
was used to calculate exchange correlation energy. The kinetic energy
cutoff of plane-wave basis was set to 60~Ry and a 6 \texttimes{}
6 \texttimes{} 6 Monkhorst-Pack grid \citep{Monkhorst1976} was used
to perform Brillouin-zone integration.

\begin{figure}
\centering{} \includegraphics[width=0.8\columnwidth]{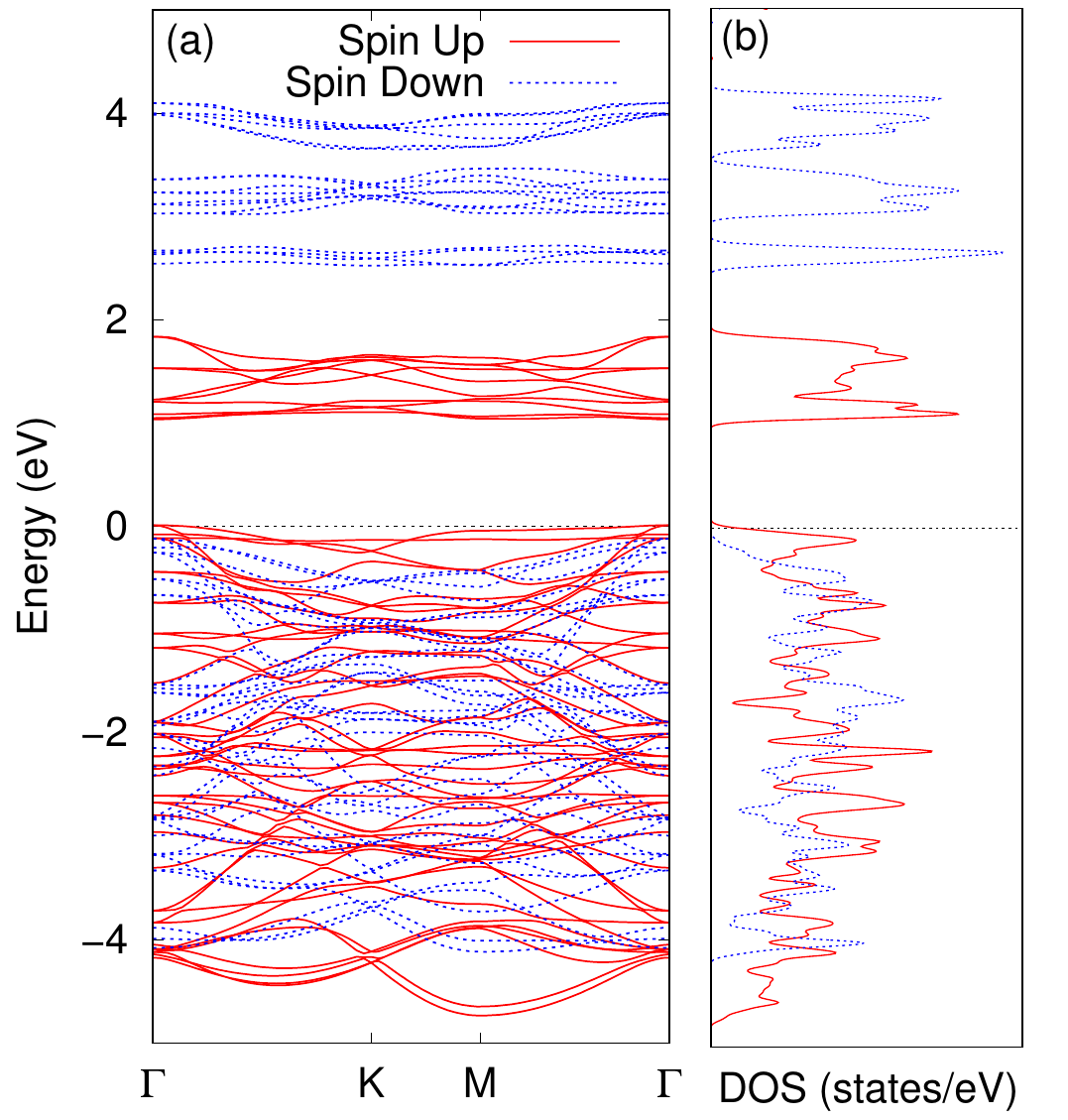}\caption{\label{fig:TheoryPlots}Calculated (a) electronic band structure and
(b) density of states of a VI$_{3}$ single crystal (room temperature
structure).}
\end{figure}
\begin{figure}
\centering{} \includegraphics[width=1\columnwidth]{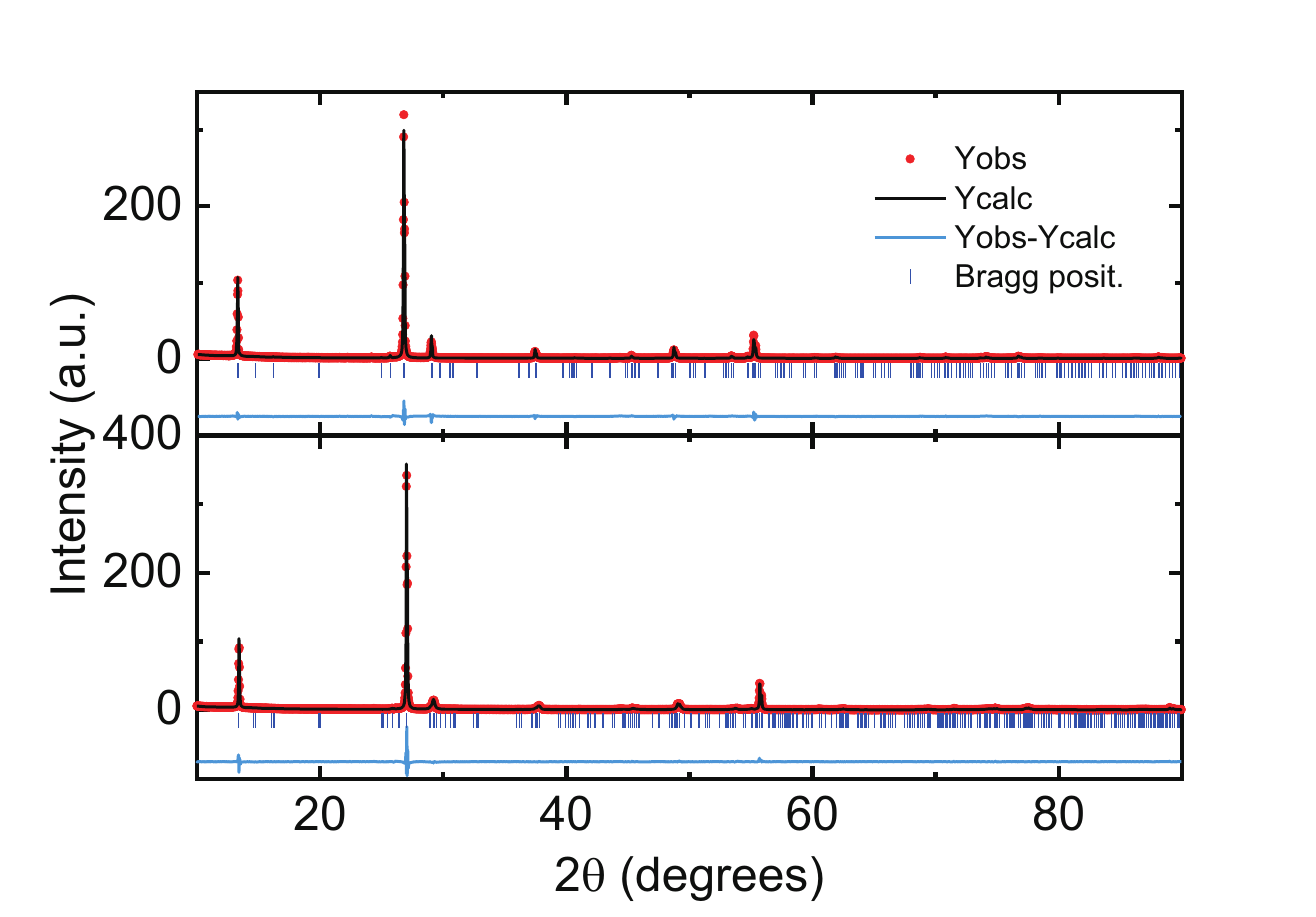}\caption{\label{fig:RefinementPlots}Rietveld refinement plots for the high-temperature
(upper) and low-temperature (lower) structures.}
\end{figure}

\begin{figure}
\centering{} \includegraphics[width=1\columnwidth]{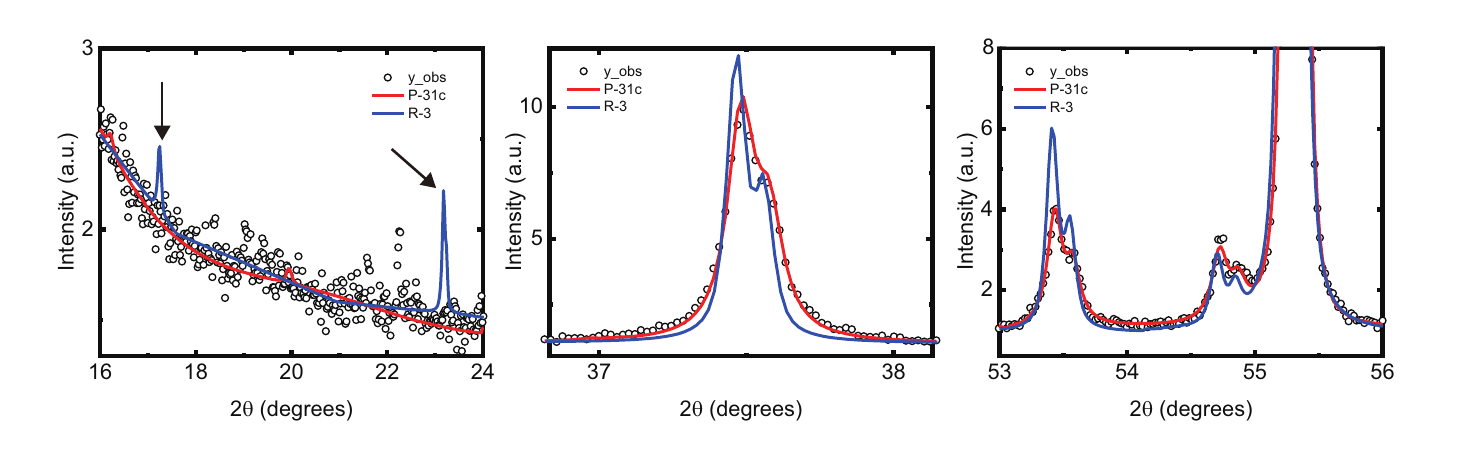}\caption{\label{fig:StructureDetails}Details of refinement fits of the $R-3$
and $P-31c$ structures at room temperature. The $P-31c$ structure
can be seen to fit several features markedly better. }
\end{figure}
\begin{figure}
\centering{} \includegraphics[width=0.8\columnwidth]{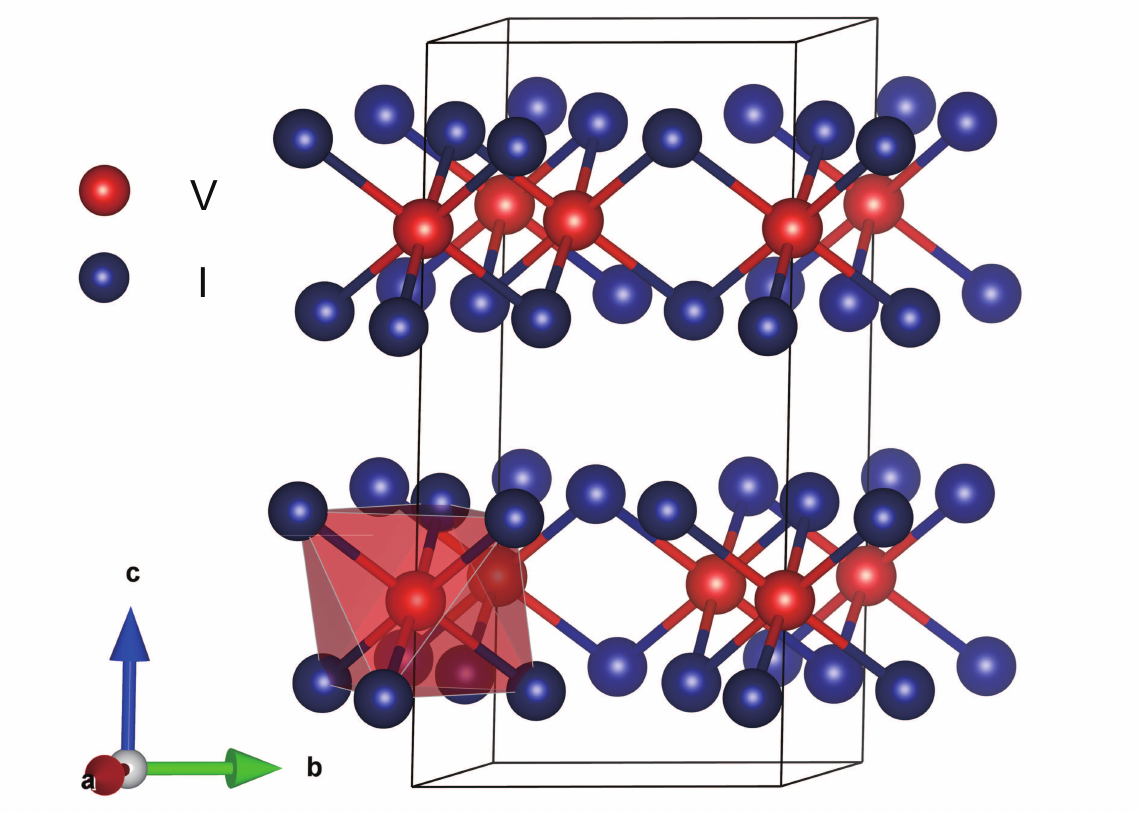}\caption{\label{fig:RoomTStructure}High-temperature crystal structure at 300~K.
Vanadium sites are displayed in red, iodine in blue.}
\end{figure}
\begin{figure}
\centering{} \includegraphics[width=0.8\columnwidth]{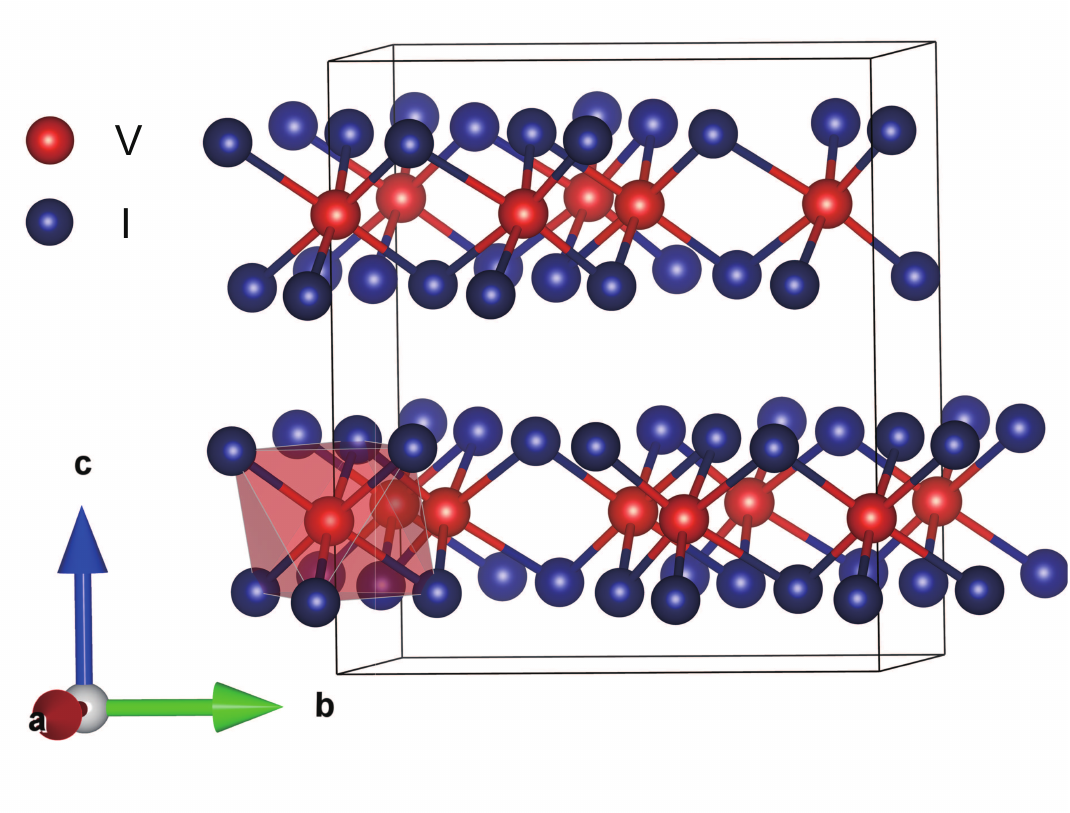}\caption{\label{fig:LowTStructure}Low-temperature crystal structure at 40~K.
Vanadium sites are displayed in red, iodine in blue.}
\end{figure}
\begin{figure}
\centering{} \includegraphics[width=1\columnwidth]{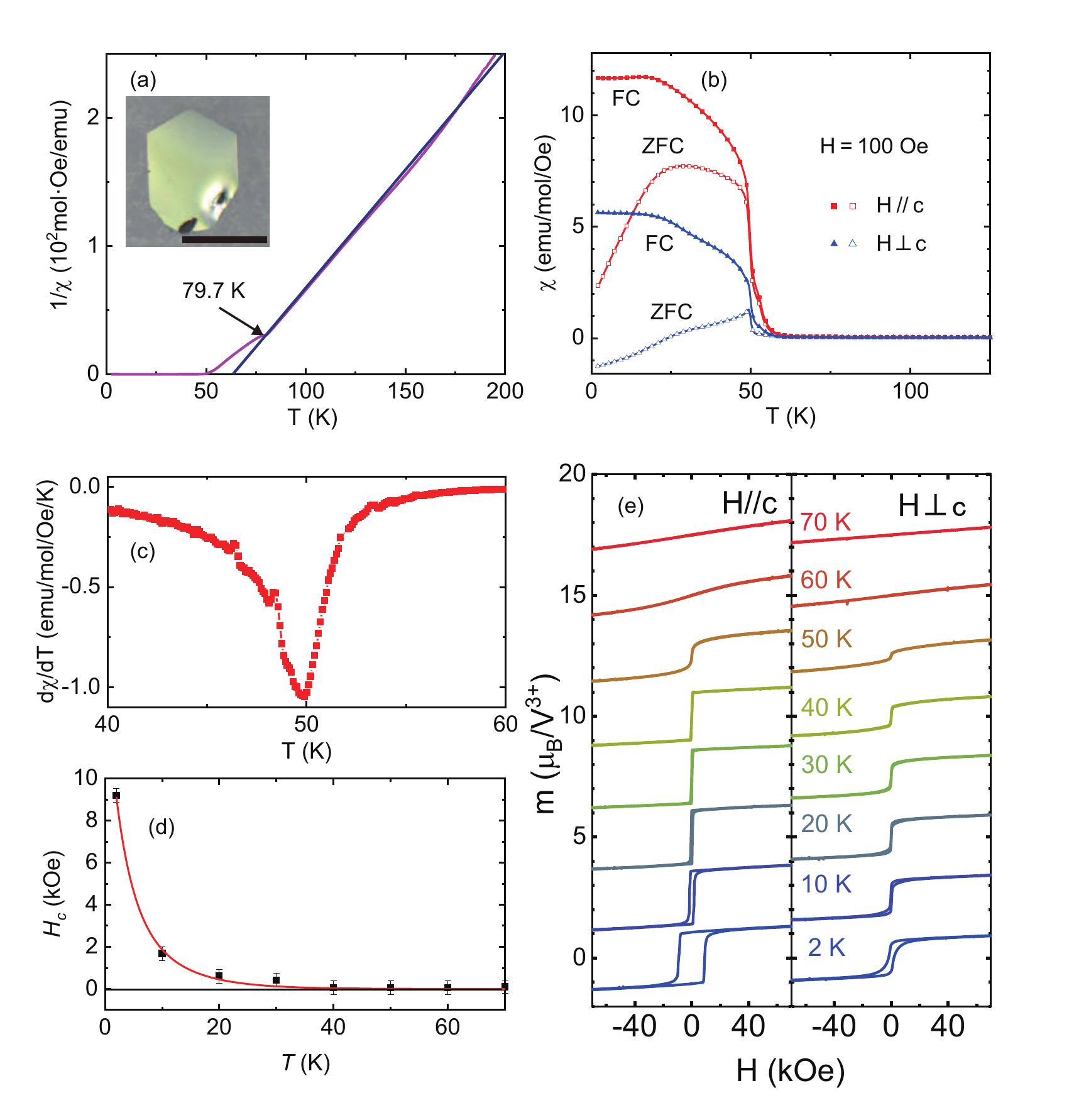}\caption{\label{fig:MagnetisationDetails}Further detail of the magnetization
data. (a) ZFC inverse susceptibility, showing good Curie-Weiss dependence
and the kink at T$\mathrm{_{s}}$. Inset shows a typical crystal,
with 1~mm scale bar. (b) Comparison of ZFC and FC susceptibility
for the two crystal orientations. (c) The gradient of the susceptibility
used to extract T$\mathrm{_{c}}$. (d) Temperature dependence of the
coercive field, and an exponential guide to the eye. (e) magnetic
hysteresis loops at more temperatures than shown in the main text,
offset for clarity. }
\end{figure}
\begin{figure}
\centering{} \includegraphics[width=1\columnwidth]{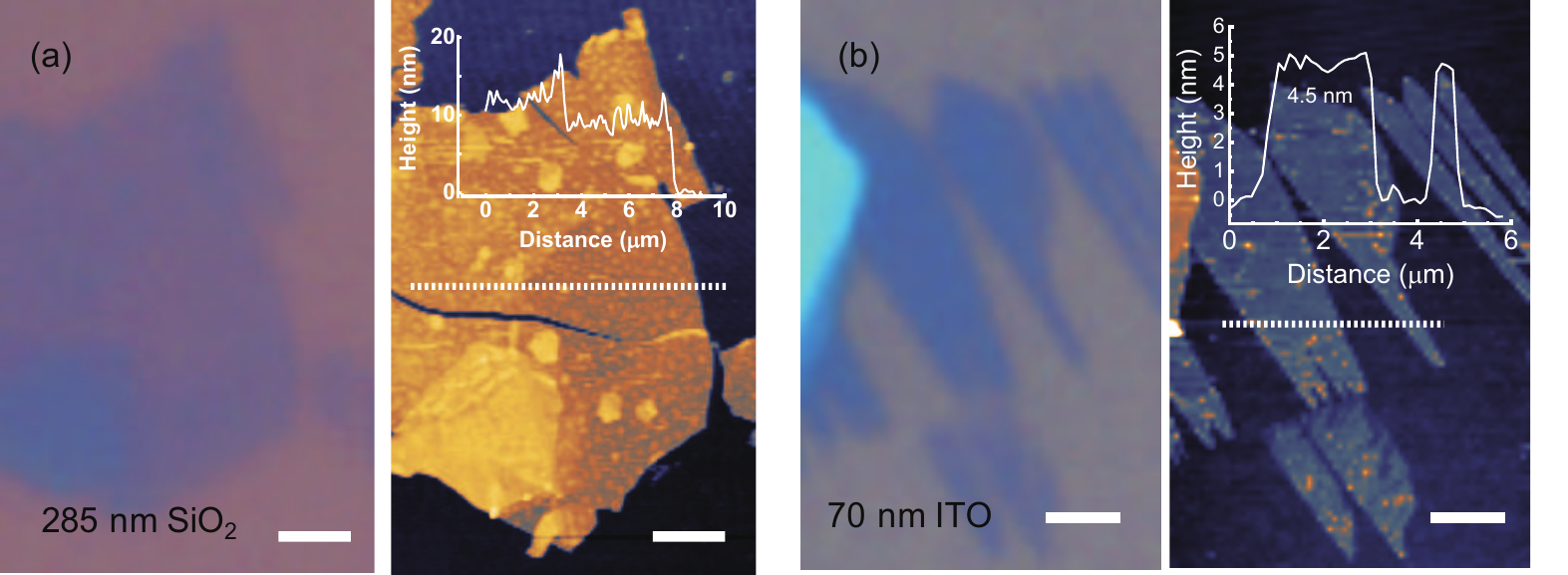}\caption{\label{fig:AFM}Optical microscope image and non-contact AFM image
of exfoliated VI$_{3}$ flakes on a (a) silicon / silicon oxide and
(b) indium tin oxide substrate. Scale bars are 2~$\mu$m. ITO gives
clearer optical contrast for few-layer samples - the pictured data
are 5~nm thick - approximately 4 unit cells. }
\end{figure}
\begin{table*}
\begin{centering}
\begin{tabular}{>{\centering}m{0.14\textwidth}>{\centering}m{0.14\textwidth}>{\centering}m{0.14\textwidth}>{\centering}m{0.14\textwidth}>{\centering}m{0.16\textwidth}>{\centering}m{0.16\textwidth}}
High-T structure & 300~K & $P-31c$  &  & $R$ = 7.37\% & $wR$ = 10.1\%\tabularnewline
\addlinespace[0.0075\textheight]
$a$ = 6.8987(10)~\r{A}& $c$ = 13.2897(1)~\r{A}&  &  & V = 547.74(10)~\r{A}$^{3}$ & \textgreek{r} = 5.236(1)~g.cm$^{-3}$\tabularnewline\addlinespace[0.0075\textheight]
\midrule 
 & $x$ & $y$ & $z$ & Occ & $U_{iso}$\tabularnewline
\midrule
V(2a) & 0 & 0 &  0.25 & 1 & 0.038(5)\tabularnewline
V(2c) & 0.33333 & 0.66667 & 0.25 & 1 & 0.265(31) \tabularnewline
I(12i) & 0.326(2)  & 0.335(2) & 0.3703(1)  & 1 & 0.041(1) \tabularnewline
\bottomrule
\end{tabular}
\par\end{centering}
\vspace{0.8cm}
\begin{centering}
\begin{tabular}{>{\centering}m{0.14\textwidth}>{\centering}m{0.14\textwidth}>{\centering}m{0.14\textwidth}>{\centering}m{0.14\textwidth}>{\centering}m{0.16\textwidth}>{\centering}m{0.16\textwidth}}
Low-T structure & 40~K & $C2/c$ &  & $R$ = 6.94\% & $wR$ = 9.34\%\tabularnewline
\addlinespace[0.0075\textheight]
$a$ = 6.9354(3)~\r{A}& $b$ = 11.9069(5)~\r{A}& $c$ = 13.1865(1)~\r{A}& \textgreek{b} = 90.403(2)~\degree& V = 1088.90(7)~\r{A}$^{3}$ & \textgreek{r} = 5.266(1)~g.cm$^{-3}$\tabularnewline\addlinespace[0.0075\textheight]
\midrule
 & $x$ & $y$ & $z$ & Occ & $U_{iso}$\tabularnewline
\midrule
V(4e) & 0 & 0.010(6) & 0.75 & 1 & 0.03900 \tabularnewline
V(4e) & 0 & 0.356(5) & 0.75 & 1 & 0.09000 \tabularnewline
I(8f) & 0.834(2) & 0.168 & 0.63 & 1 & 0.04740\tabularnewline
I(8f) & 0.835 & 0.497 & 0.627 & 1 & 0.04740\tabularnewline
I(8f) & 0.33 & 0.334  & 0.6271(3) & 1 & 0.04740\tabularnewline
\bottomrule
\end{tabular}
\par\end{centering}
\caption{\label{tab:refinementParamsTable}Refinement parameters for the high-
and low-temperature phases.}
\end{table*}
\end{document}